# AI Narrative Breakdown. A Critical Assessment of Power and Promise

Rainer Rehak
Weizenbaum Institute for the Networked Society
Berlin, Germany
Berlin Social Science Center (WZB)
Berlin, Germany
rainer.rehak@weizenbaum-institut.de

## Abstract
This article sets off for an exploration of the still evolving discourse surrounding artificial intelligence (AI) in the wake of the release of ChatGPT. It scrutinizes the pervasive narratives that are shaping the societal engagement with AI, spotlighting key themes such as agency and decision-making, autonomy, truthfulness, knowledge processing, prediction, general purpose, neutrality and objectivity, apolitical optimization, sustainability game-changer, democratization, mass unemployment, and the dualistic portrayal of AI as either a harbinger of societal utopia or dystopia. Those narratives are analysed critically based on insights from critical computer science, critical data and algorithm studies, from STS, data protection theory, as well as from the philosophy of mind and semiotics. To properly analyse the narratives presented, the article first delves into a historical and technical contextualisation of the AI discourse itself. The article then introduces the notion of "Zeitgeist AI" to critique the imprecise and misleading application of the term "AI" across various societal sectors. Then, by discussing common narratives with nuance, the article contextualises and challenges often assumed socio-political implications of AI, uncovering in detail and with examples the inherent political, power infused and value-laden decisions within all AI applications. Concluding with a call for a more grounded engagement with AI, the article carves out acute problems ignored by the narratives discussed and proposes new narratives recognizing AI as a human-directed tool necessarily subject to societal governance.

## 1 Introduction

With the release of the interactive chatbot ChatGPT by the US-American company OpenAI in November 2022, an application of artificial intelligence (AI) was introduced for the first time that delivers impressive results even for laypersons and that can be used as a text-based web application by anyone without prerequisites. This combination made it possible for the media world as well as interested parties from politics, business, and civil society – including private individuals – to immediately try out this kind of technology. Discussions on current capabilities and possible developments of AI thus became broadly relatable within society, and the topic became an urgent societal issue virtually overnight.

Although there are many types of artificial intelligence, the discussions about the possibilities and limits of ChatGPT and similar programs reveal some recurring narratives that currently shape societal engagement with AI in general. Even more importantly, those narratives are not only part of the contemporary AI discourse, but narratively conceptualizing AI shapes the very discourse space itself, including its (im)possible forms of epistemic questions, practical judgments and general notions of potential. Concretely, narratives play a crucial role in (co-)imagining futures [58] and eventually have manifold real-world implications, e.g. on design, use and regulation of AI [5]. The endeavor of this article is therefore to take a closer look at the most prevalent AI narratives, critically engage with them and, if necessary, dissect and deconstruct them [110].

The article begins with a brief historical and technical overview of the AI discourses relevant to analyse popular AI narratives. It then moves on to describe and critically reflect on narratives such as agency and decision-making, autonomy, truthfulness, knowledge processing, prediction, general purpose, neutrality and objectivity, apolitical optimization, sustainability game-changer, democratization, mass unemployment, and societal utopias/dystopias through AI. Throughout the text, abstract conclusions are illustrated by insightful concrete examples. The article concludes with an overall summary of the missing aspects of the current societal discourse on artificial intelligence and proposes constructive alternative discourses looking beyond misleading narratives.

### 1.1 The AI Discourse

The term artificial intelligence was introduced in 1955 by U.S. computer scientists including John McCarthy and Marvin Minsky to present a research program "simulating" intelligence at the Dartmouth Conference and to solicit research funds with it [77]. While the basic idea of intelligent computers is older and more related to, for example, communication with humans [117], computer intelligence was now described as performing any task that had previously been exclusive to humans [77].

The technical development of artificial intelligence has always been accompanied by a narrative discourse on the possibilities and limits of the respective systems. In 1966, the computer scientist Joseph Weizenbaum developed a simple chat program named ELIZA – today, we would call it a symbolic AI application – that was meant to demonstrate how easy it is to simulate human understanding [125]. The program was designed to mimic a psychotherapist by inserting terms from the users' text inputs into pre-configured question constructions ("Do you enjoy being X?") or by providing generic responses ("Please go on."). It astonished Weizenbaum that the chat program was actually attributed empathy and personality, within and beyond the professional world. These reactions made him one of the most vehement critics of, how he called it, mythical technology belief, exaggerated expectations of automation, and the tendency to anthropomorphize computers ([126]; more recent [7, 95]). Since then, the "ELIZA effect" describes the inclination to attribute human characteristics such as intelligence or consciousness to AI systems if they mimic typical human expressions, appearances, or behaviors just well enough. Concrete and usable AI breakthroughs, however, were missing in the 1960s and 1970s [73].

The economic consequences of AI were also discussed fifty years ago. The philosopher Hubert Dreyfus wrote critically about the AI hype at the time in 1972: "Every day, we read that digital computers play chess, translate languages, recognize patterns, and will soon be able to take over our jobs" [31] – a forecast that was regularly heard even then and continues to play a significant role in more recent debates [11, 43]. It is certainly true that the field of artificial intelligence has now achieved such great successes that everyone working with computers also regularly uses AI systems [91], whether indirectly in search engines or directly for translating as well as text and image generation.

This very brief summary shows that the 70-year history of this technology has been characterized from the beginning by grand narratives of an allegedly imminent AI revolution of thinking machines [114] that would change everything forever [68], often with an apocalyptic undertone [52, 81].

To provide structure in this sometimes confusing mix of contemporary diagnoses and futuristic predictions, the following sections will outline and discuss currently prominent narratives about artificial intelligence. On this basis, appropriate discussions about the socio-political implications of AI systems can then be initiated – covering conceptualisations, potentials, dangers, regulation, and the very material shaping of these technologies.

## 2 Two Types of AI and a Discourse Tool

To analyze narratives and hence understand better the conditions under which artificial intelligence is currently negotiated in society, it is necessary to conceptually distinguish between two types of AI systems. This established differentiation is regularly picked up in discourses on artificial intelligence, yet the terms are often used inaccurately and even mixed, although they have different characteristics and implications [95].

Firstly, there are domain-specific AI systems (artificial narrow intelligence, ANI), sometimes also called weak AI, which are designed and developed for a specific task area. They are optimized for certain processes but are only useful within this context [22, 78]. For example, a program for the energy-efficient air conditioning of data centers is pre-configured with selected load data from the past so that it economically regulates the use of cooling units according to need; however, it could not be used to curate a music playlist. Likewise, a chess program cannot generate images, nor can an image generator AI perform explicit mathematical regression analyses. Therefore, anthropomorphizing terms like "self-learning" are inappropriate [95]. All currently existing AI systems undoubtedly fall into the category of domain-specific artificial intelligence [101], from modern Go computers to image recognition or translation software, to the large language models (LLMs) such as OpenAI's GPT, Google's PaLM/Bard, BAAI's WuDao, or Meta's LlaMa.

Secondly, there's the concept of universal AI systems (artificial general intelligence, AGI), sometimes called strong AI, which are purportedly truly capable of autonomous learning and abstract thinking, and which possess creativity, motivation, consciousness, and emotions [115, 117]. In some conceptions, these AGIs even develop superhuman abilities [68, 81]. However, to date, there is neither a functioning AGI nor reliable evidence that such a system can be developed with current computer architectures at all [72]. Although there have been research projects and competitions in this direction for decades – such as the well-known Loebner Prize, which uses a version of the Turing Test – the results have always been sobering. Moreover, the widely discussed AGI milestones, i.e., which technical problems need to be solved for it, have regularly shifted. Formerly, beating the world chess champion was the goal; now, automatic text generation is the benchmark. Many experts even fundamentally doubt the possibility of general artificial intelligence in the form of a digital computer [31, 39, 99]. Nevertheless, AGI is still often discussed, which might be related to the business motives of economic actors [38], but also to the central role such systems play in science fiction works, see for example Samantha in 'HER', Data in 'Star Trek', HAL 9000 in '2001', Ava in 'Ex Machina', C3PO in 'Star Wars', Bishop in 'Aliens', T-800 in 'Terminator', or even the Maschinenmensch in 'Metropolis' [50].

To characterize the regularly (too) vague use of the term artificial intelligence in societal discourse, I propose the term "Zeitgeist AI" [96]. "Zeitgeist AI" can mean anything from big data, algorithms, apps and statistics to software, robots, digital technology, all the way to the internet and digitalization in general. In political discourse, too, there is often a blanket reference to "artificial intelligence," regardless of whether it concerns self-driving cars, robotic dogs, automated decision-making systems, climate models, automated public job placement systems, table reservation systems, or smart traffic control systems; at times, even traditional computer science products are labeled as such. Everything counts as "AI", even if there is little to no artificial intelligence in all of it. Calling anything digital "AI" prevents any meaningful discourse, but surely signals modernity. For example, a German report from 2022 on the situation of AI use in public administration states that "often projects are described as AI-based, but in fact, they use conventional ICT applications." [13]. But this vague understanding is also found in science and business (science: [56]; business: [122]). But to enable a fruitful societal engagement with artificial intelligence, the cloudy use of "Zeitgeist AI" must be called out, the real respective meanings be explicated and thus unnecessary confusion resolved (as positive example see [41]).

## 3 Discussion of Current AI Narratives

Against this background, we can now outline and discuss common narratives about artificial intelligence by asking on what assumptions do they rest, to what extent do they correspond to the current state of research, and what socio-political implications do they have. Often, these narratives overlap and intermingle, so the following presentation makes a distinction by key notions, serving more as a spotlight analysis of discourses on AI rather than a systematic mapping of its subject.

While the analysis of and reflection on AI narratives might seem like a theoretical finger exercise, it is crucial to understand, that common narratives shape the space for negotiating technology, or to re-interpret David Beer "it is also the very concept of [AI] itself that shapes decisions, judgments and notions of value" [6]. Therefore narratives of AI will eventually unfold concrete material consequences for academia, civil society, politics, and society at large. Hence, critical analyses of such narratives are useful and necessary.

This article builds on previous academic work on digital and AI narratives by scholars such as Alan Turing (1950) [117], Mortimer Taube (1961) [114], Hubert L. Dreyfus (1972) [31], Joseph Weizenbaum (1976) [126], Vincent Mosco (2005) [80], Evgeny Morozov (2014) [79], Lucy Suchman (2023) [110] and others [5, 22, 28, 40, 49, 53, 74, 95, 96, 103, 107, 109], and extends it by analysing more recent AI applications and use cases, and strong new narratives like AI as sustainability game-changer, while maintaining a close coupling of concrete technical, historical and philosophical discourses. Primarily fictional AI narratives [19], however, are not in scope of this article.

The analyses and conclusions in this article are based on insights from critical computer science, critical data and algorithm studies, STS, data protection theory, as well as from the philosophy of mind and from semiotics.

### 3.1 Agency

AI systems are often attributed with the ability to "act" or "decide" something [37]. Such attributions implicitly assume that AI systems are capable of agency (as opposed to mere behavior) and are not just (complex) tools that implement an externally set purpose. However, agency implies having one's own intention, an internal representation of the situation, a potentially alternative outcome of action, moreover, a moment of conscious decision, and ultimately also responsibility [95]. AI systems, however, get their "intentions" set externally and behave deterministically. If the results are not deterministic, as with ChatGPT, it is due to deliberately built-in random parameters and thus decisions made during development, not an alleged independent agency of the AI [131]. Even if emergent behaviour of digital systems could be observed, like in the famous high frequency trading crashes in 2012, this is due to complex randomness or error, but intention and an internal representation are not present at all. As long as there is no AGI, the outcomes of all AI systems are primarily the result of design decisions, including random factors or errors. If the deployed systems do not fulfill their purposes, manufacturers and operators modify them, make adjustments, and correct issues, which further emphasizes the tool-like nature of such systems. The fact that applications of artificial intelligence resemble large technical systems rather than simple tools does not change their mischaracterisation as independent agents, which only diverts from the actually responsible organizations (companies, authorities, etc.) that program or purchase and deploy these systems for their own purposes. Therefore, these systems have no agency, at least not in a meaningful sense [15].

### 3.2 Autonomy

Closely connected to the concept of agency is the narrative of autonomy of AI systems. This attribution got much visibility in the recent years through the discourse around so-called autonomous drones and autonomous driving, although the notion of autonomous systems has been critically discussed within computer science since the 1990s [46], e.g. in applications for trains, air planes, and industrial production [27]. While the concept of autonomy originally stems from the political notion of city-state self-determination (Greek: αὐτός/autós: self, νόμος/nómos: govern) in Greek political philosophy [90], nowadays it is usually understood as being able to make informed, uncoerced decisions as individuals or groups, so its connotation of independence reaches far beyond mere agency. However, similar to other terms in the AI discourse, autonomy initially meant something else in the technical and AI discourse [95]: Coming from highly static approaches to automation like conveyor belts and robot assembly lines, where goals and means are precisely defined, i.e. process units are clearly specified and implemented, the term "autonomous" technically just meant taking steps towards more flexibility regarding the means. Therefore, not unlike intelligence itself, the term autonomy is not universally defined and definitions primarily exist only for specific use cases in heavily standardized environments like in factories, air travel, highways or railway use [55]. The proper technical term, contextual autonomous capability, hints to its contextual limitation. In the mid 2010s, autonomous driving, especially within urban areas, became a big topic and PR stage for tech giants which also fueled triumphant AI narratives. However, AI only plays a role in autonomous driving in terms of object detection or usability improvements, e.g. steering performance, not regarding high-level functions [95] or other aspects we usually associate with autonomy. To escape the mythical ascriptions, "autonomous driving" should therefore instead be called automated driving. The same is generally true for all "autonomous" systems: the goals and rules have to be externally determined, then the system dynamically pursues the given tasks [15].

### 3.3 Truthfulness

Current AI systems can neither lie nor con. Naturally, their statements can be false (and often are, [40]), but a lie implies knowledge of the truth and a deliberate departure from it. Similarly, conning implies knowledge of one's own identity and a deliberate pretense of being someone else. AI systems cannot make truthful claims, since they output exactly what their model architecture and data dictate. Therefore, they also cannot carelessly talk nonsense, manipulate strategically, or distract consciously, as all this would require at least an internal state, motivation, a model of the mind, and a model of the other. As described above, AI systems consist of complex but purely formal processes. The deploying organizations, on the other hand, do have interests and potentially also knowledge of the truth and their own identity. Hence, corresponding claims must be made against the operating companies and organizations.

### 3.4 Knowledge processing

Current text-based AI applications like ChatGPT rely on language models that calculate multidimensional vector spaces from vast amounts of input texts, where the "distance" between words and types of words is stored; this is the language model. In normal use, texts are generated by the model determining a very likely next word based on a user input (a "prompt"). Then, the prompt plus the first result word become the internal input for the next word. This process is repeated multiple times. Eventually, the output ("response") of the chatbot is composed [131]. The outputs are therefore formal-mathematical sequences of character strings, statistically recombined from the input texts, yet, their meaning – unlike the less powerful symbol-based knowledge models – technically cannot matter to the system [7]. Consequently, truth or accuracy are not relevant criteria for the responses and cannot easily become so. LLMs are formal language models, not knowledge models. The results of such AI systems therefore (re)produce the formal relationships between words in the input texts, which is precisely their function. The results are sometimes impressive, but the errors are equally obscure, such as when meaningless platitudes or pure "fantasy facts", often wrongly called hallucinations [32], are being generated. Strictly speaking, language models cannot "make mistakes", the sequences of words in the input texts are just as they are (cf. [4]). The style – the form – of the generated texts, however, is usually impeccable, as the texts the model was trained on are usually very well-styled. Therefore, it makes little sense to claim that ChatGPT can "even" produce texts in the style of poems, academic articles, news reports, or manuals, because this mimicking is precisely what language models are optimized for, and nothing else is possible with this technology [74]. Mathematically speaking, the results are variations of a statistical average text within the model relative to the specific query. A fitting analogy would be a stochastic parrot that reproduces sequences of words according to statistical rules [7]. We are dealing neither with real language competence nor with understanding or knowledge in either case.

### 3.5 Prediction

AI systems are statistical systems and can therefore detect formal patterns and dependencies of variables in existing datasets, be it weather, business, or behavioral data. These patterns necessarily always refer to the past but can be statistically evaluated and mathematically projected into the future. The extent to which this projection then qualifies as a prediction in the actual sense depends on the subject area. If its data and dependencies are subject to physical laws, like weather data, this can work. However, if it involves social data, the calculations only generate a true prediction, if the future will be exactly as the (data) past was – this assumption of social laws is highly problematic from the perspective of scientific theory and social science alike, even if the past data were complete [36]. The rise and fall of predictive policing illustrate the problem well, as it was assumed that crime could be geographically predicted in a similar way to earthquakes. However, the underlying processes are very different – societal processes, unlike physical ones, do not follow fixed laws. In practice, the results of predictive policing software were of little use – in some cases, the predictions were accurate in less than one percent of cases, and in other cases, they even reproduced racism, by disproportionately indicating crimes in black neighbourhoods [47, 102].

### 3.6 General purpose

Oftentimes, powerful AI systems like LLMs are called general purpose systems (e.g. [89]). Yet, none of these systems are "general purpose" in any meaningful way. It is of course possible to apply those systems to all kinds of contexts in all kinds of sectors because there will always be a concrete response to a given prompt from the system, but it always takes a lot of contextual knowledge to determine if the response is actually useful. In some cases the text results have indeed been seemingly satisfactory (e.g. [61]) but in many others they were rather dysfunctional [116] or even utterly catastrophic (e.g. [4]). The usefullness really depends on the task and in many cases the outputs are merely well-sounding commonplaces or fabrications and falsifications [32]. Those tools could therefore be useful for entertainment purposes, for inspirational art creation, for programming assistance, and yes, for mass-producing spam [18], but this is a far cry from being "general purpose". Seen in this light, the general purpose narrative turns out to be a greatly misleading marketing term.

### 3.7 Neutrality and Objectivity

Digital and data-based systems are often attributed neutrality and objectivity, and this is also the case with AI systems. However, in the process of designing the system, e.g. the information processing, the data used, the outputs created etc., there are always degrees of freedom and therefore decisions to make, which fundamentally shape the resulting system. Any explicit digital system is the result of modeling what it is supposed to do. For example, if I want to program a digital ride hailing platform, I should model the car, the driver, the driver's rating, the prices and so on. However, I would probably not model the happiness of the driver, the type of tyres or the type of window glass used, because this is usually not necessary for organising ride hailing. Yet, the former one would be of great interest for worker's unions and the latter two for car repair shops, so those are not necessary only from a specific point of view [35]. The same questions can be asked about how and why social network companies model social connections and interactions the way they do it or how public agencies model their citizens and so on. Different purposes require different systems and even for one purpose, regularly several modelings are possible. Why not create one huge model of everything? Well, there is no total, consistent and objective model of the world, hence all digital systems and their data are the result of pragmatic reduction [108]. However, pragmatic reduction is a necessarily subjective and partisan notion. In effect, modeling "all aspects" of a given situation is not only not feasible, it is simply not possible. Therefore, coming from statistics, the ML community adopted the telling aphorism "All models are wrong, but some are useful" [104]. To summarize: Specific actors design systems according to their particular purposes.

Depending on this purpose, they also select the data sources for the system, which in turn is crucial for the system's results. For similar reasons as explained above, concrete data itself is also neither objective nor neutral [62]. For example, it is possible to install particle sensors in a city to measure air quality, but the decision about the particle sizes, the sensor height and locations for these sensors (parks, streets, kindergartens, mountains etc.) depends on the actors and their purposes. And because it is, again, impossible to measure "everything", any data at hand already is the result of many decisions. In other words "raw data is an oxymoron" [44]. Of course there are use cases with less contested data decisions like weather or astronomical data, but as soon as data is supposed to represent social or societal aspects or even touches them, negotiating the semantics begins. For illustration here is another example: If predictive policing applications would base their calculations of future crime hot spots not only on the types and locations of past crimes, but also on the damage value, the prediction systems would primarily identify the main business districts and financial trading places of cities as future crime hot spots. Adding or removing damage value for such calculations is a subjective decision and such decisions are necessary for all data-based systems – including AI systems, which is why none can be considered "neutral" or "objective" [33, 93].

Currently, a lot of AI research in academic publications and conferences deals with questions of fairness of AI systems and how to prevent discrimination. However, these approaches often fail because fairness is regularly treated as correcting an erroneous bias from an objectively correct result. But an automated credit granting system based on a person's income, for example, can currently only be mathematically correct or actually fair: either it is correct but then reproduces the gender pay gap, or it is fair but mathematically incorrect [60]. Hence, fairness is only to a small degree a problem to be solved technically. Furthermore, there is not one single consistent understanding of fairness and many understandings even contradict each other, also depending on the use case and the application context [120]. In effect, the problem if a given system can be considered non-discriminatory and fair always has to start with collective negotiation processes [36]. Not explicitly doing this does not imply its objectivity, but rather the existence of lots of hidden assumptions, many of which almost certainly benefit the designing actor.

## 3.8 Apolitical Optimization

At times, huge societal and political issues are portrayed as tasks essentially solvable by AI [3, 20]. This claim includes, for example, introducing an automated unconditional basic income [84], global road safety [45] or ending world hunger [17]. According to the narrative, this approach appears possible, even promising, because AI is able to generate "apolitical" and "unideological" solutions for wicked societal problems. How problematic and wrong this narrative is, can be illustrated with three examples:

**A)** Using AI for predictive maintenance can help to reduce unexpected down-times of mechanical machinery (goal) by continuously measuring certain physical properties like vibrations or sound and based on that, detecting anomalies hinting towards components likely to fail in the near future. This allows for conveniently choosing a timely repair window and also preventing uncontrolled damage to connected parts (strategy). This application of AI is possible without much need for social negotiation, because both, goal and strategy are rather uncontroversial.

**B)** Using AI for improving learning success of pupils would surely be an agreeable goal, but doing so by deploying dynamic learning journeys, while processing extensive data on behaviour, learning developments and social connections touches upon many fundamentally controversial issues like pedagogical understandings (including reflecting on the teachers' role) or data protection issues. In this case, the goal might be uncontroversial, but the optimal achievement strategies to be determined by AI are highly dependent on the understanding of acceptable approaches. So even if there is agreement regarding a given goal, there are usually countless strategies to achieve it and the choice of intermediate steps and the weighting of relevant factors are often not technical questions [88, 93].

**C)** Using AI to make better political or managerial decisions seems like a sensible goal at first, but taking a closer look this turns out to be just a rhetorical dodge, because calling for "better X" nicely evades the question what "better" concretely means. More often than not, there is no universally accepted understanding of betterness, especially in social or societal contexts [93]. But in order to actually optimize a situation or a process, it is necessary to define what exactly is being sought for as a goal (or more technical: as objective), what should then be considered improvements in the specific subject area and what approaches or degrees of freedom are present to move forward. Hence, when we ask about the specific differences between good political or managerial decisions in contrast to bad ones, we will get a plethora of different answers and often even contradictory ones. So, finding a proper objective function in order to employ an AI system requires prior social or even societal negotiation, which is, by definition, a political process. Only after those details have been thought out and defined, AI systems can be applied to improve and optimise.

Those three examples illustrate that the more complex the task is, the more design decisions have to be pre-negotiated before even applying AI. Thus, the inherent contradiction of the apolitical optimization narrative lies in the assumption that AI systems are able to automatically produce precisely those conditions as a technical result, that are necessary beforehand to meaningfully apply those systems in the first place. Arguing for apolitical optimizations using AI is therefore circular reasoning. A fitting term here could be anti-political solutionism, if we want to adapt it from the critical big data discourse [93].

Keeping this in mind sensitizes us regarding the other cases mentioned above, like the fight against world hunger. What should the strategic options to be analysed by AI comprise: Canceling debts, the change of international economic treaties, the free (re)distribution of food, making available certain technologies globally, or entirely different measures? Who should finance that, who should not? All these are political questions that AI systems cannot help to solve. At best, AI can help calculate example scenarios and possibly facilitate implementation, but selecting and modeling the relevant conditions is, again, a task that primarily requires collective, i.e., political, deliberation and decisions.

### 3.9 Sustainability game-changer

One specific recent narrative closely related to the apolitical optimization one concerns eventually averting climate change and globally reaching sustainability through AI [3, 85, 118, 119]. This narrative gets special attention here, because the intersection of sustainability and digitalisation is currently very present in public and academic discourses, and is, in its extreme form, a showcase of magical thinking comparable only to the one in the blockchain hype debate at that time [94].

The narrative, that AI is the one key breakthrough technology bringing solutions to the Earth's (un)sustainability problem, is based on two sub-narratives. On the one hand, there is the apolitical optimization narrative discussed above, where the claim is basically, that AI solutions are above politics and can therefore produce objectively good solutions agreeable by everyone. On the other hand there is the information problem narrative stating that a core barrier to achieve sustainability is a lack of information and knowledge, whereas AI systems can close the relevant data gaps and finally allow for sustainability innovations beyond imagination. I will expand both aspects.

I have already dealt with the apolitical optimization narrative, yet I want to give two more sustainability-related examples in the same style as above to make more precise my critique, that the sustainability game-changer narrative is fundamentally misleading:

**A)** Using AI can help to reduce the energy consumption of data centers (goal) by calculating energy-efficient cooling cycles based on past and present computation load data (strategy) [34]. This is easily possible here, since the goal and the strategy are largely uncontroversial. However, this does not prevent the construction of additional data centers that now become profitable through the reduced cooling cost, so that the overall energy consumption still rises. This commonly called rebound effect [42, 128] shows the large limitations of efficiency based sustainability approaches. And as long as we still do not have 100 % renewable energy (RE) in the power grids, additional "green" data centers running exclusively on RE essentially take away this limited resource from the grid, apart from other material and social implications in production and disposal of data center equipment. This example is not meant to downplay the technical advantages of AI applications, but to underline their limited overall effectiveness [66, 70].

**B)** In the field of urban planning and development, AI applications are said to greatly help with designing more healthy, liveable and eventually sustainable cities (and non-cities). This goal, as above, sounds desirable, but assuming AI excels in those tasks has crucial shortcomings regarding the actual meaning of a "healthy, liveable and sustainable" city. Not only can AI not automatically determine priorities like deciding between strengthening individual traffic and strengthening public transport, it can, in principle, also not set the optimisation criteria regarding balancing the interests of car owners with the safety of cyclists, the availability of inner city parking with the existence of pedestrian areas or parks, even deciding to take the long term health effects of biking into account in comparison with the present convenience (and emissions) of cars in smart parking systems. In what way would an AI system calculating a car-free bike city as "optimal" be perceived as more objective than a politician saying the same? In the case of the given AI result we would immediately have a political discussions about how the issue had been modeled within the AI system and what priorities had been programmed into it, and in effect, the actual problems would remain as unsolved as without AI.

The second sub-narrative of the assumption, that AI is crucial and essentially sufficient for reaching sustainability, is based on framing the sustainability problem as an information and transparency problem. This line of reasoning started with the big data discourse [75], but intensified further with the recent rise of AI applications. At its core lies the claim that current sustainability measures cannot be sufficient, because we lack a lot of necessary data, information and knowledge. Then, AI is said to help uncover patterns in all kinds of large and real-time data, and therefore facilitates sustainability approaches previously unheard of. While it is true, that AI has many useful applications in the sustainability area [2, 8, 25, 51, 92, 121, 123, 129, 130], it is highly misleading to assume, that the current level of knowledge is insufficient for meaningful action. Primarily focusing on getting more information is neither necessary most of the time nor does additional information automatically translate into individual or political action. Most of the time, additional data just shows existing and know problems like deforestation, overfishing, rising $CO_2$ levels, decreasing biodiversity, poverty, pollution, weather extremes, drought, hunger, inequality, conflict etc. in more detail, but without any material change [97].

Increasingly many scientists, e.g. Scientist Rebellion or Scientists for Future, become visibly frustrated and take to the streets because their sustainability related work does not have any concrete impact besides piling up more and more alarming information [57, 59, 64]. Even in academic papers one can read conclusions like this one from AI-enhanced satellite image research: "A study itself has no value if the information is not generated and shared at the right time with the right people" [111]. Statements like this clearly point not to an information gap, but to an action gap. But why do actionable insights are not put into practice? The corresponding analysis is very complex, but it centers around actors and power, around economy and politics, around equity and justice [24]. If the unsustainability issue had to be characterised by one problem, it wouldn't be the information problem, but the power problem and this is not to be solved by AI, quite the opposite. To put it in clear language: "sustainable AI is the technical solution to the climate crisis from a techno-solutionist vantage point simply reproducing the status quo. The enthusiasm for sustainable AI primarily serves hegemonic interests" [103].

To summarize it pointedly: A perfectly helpful AI information system would only be able to tell us what we already know [112]. Therefore we should not be distracted by unreasonable AI saviour narratives [97].

This narrative deconstruction explicitly leaves out the resource impact of sustainability-oriented AI systems themselves, since it already is the topic of intensive research [29, 132].

## 3.10 Democratization

Even though AI systems are complex tools currently developed by powerful corporations, it is often speculated that these tools could soon be available to everyone due to increased efficiency and other technical improvements, thereby enabling the democratization of AI [9]. What may initially sound like a desirable political concern turns out to be an economic agenda upon closer inspection. "Democratization" in this context does not mean control or co-design of such technologies by self-organized communities but merely broad access to usage for businesses [76]. The control of such large systems can hardly be democratized. While democratization might still be possible for small, highly specialized AI systems with a relatively small database, it is illusory for large AI systems like LLMs, given the immense technical, organizational (and energetic) effort required for design and production. Their development process includes (global) data collection, quality assurance, classification, storage, incremental model design, pre-configuring ("training") of the models, and does not stop at moderation and refinement of AI systems by human labor, e.g. Reinforcement Learning from Human Feedback, which is largely outsourced to the Global South [83]. Consequently, large AI systems involve a complex global supply chain [23]. Furthermore, the systems must be regularly updated, so all the mentioned steps need to be repeatedly performed. Data and systems also age over time. It is no coincidence that only financially strong players like OpenAI, Meta, BAAI, or Google can regularly announce breakthroughs in LLMs, as only they allocate the necessary capital to undertake such projects. One could argue that this kind of AI is essentially the nuclear power of the digital world [96], meaning it can only be controlled and operated centrally by powerful entities. Access is then leased to users, but the control remains with the owner. Also, this situation is not altered by the free and open AI/ML program libraries from Google and Co., by using misleading notions of openness [127] or by offering mid-size LLMs for self-operation [106], because the remaining ingredients for creation are always lacking. Or to hear it more bluntly directly from OpenAI: "As we get closer to building AI, it will make sense to start being less open. The Open in OpenAI means that everyone should benefit from the fruits of AI after its built, but it's totally OK to not share the science" [12]. Obviously, only OpenAI gets to decide what that means specifically, which somehow illustrates the whole problem.

## 3.11 Mass unemployment

Finally, let's briefly touch on the narrative that AI is a "job killer." Although this fundamental debate is as old as automation itself, it was always oversimplified; after all, the Luddites were not inherently anti-technology, but they surely were anti-exploitation. In essence, the mass unemployment narrative suggests that the use of AI would lead to the mass disappearance of jobs, with past studies even suggesting nearly half of all jobs in the USA [43] would get lost, potentially leading to mass unemployment in the foreseeable future. This narrative has at least three questionable aspects:

Firstly, it should be societally desirable that machines and AI systems take over work from humans, especially if it is repetitive, strenuous, or even dangerous [67]. The pressing questions then become, which jobs will be lost, who are the affected groups, how can they adapt, and what role should wage labor play in future societies anyway.

Secondly, studies show that jobs almost never suddenly disappear but rather change gradually. Instead of technology-induced mass unemployment, we are witnessing a structural transformation of the labor market [14, 16]. This change means that moderately and increasingly well-paid jobs are transforming into a manageable number of lucrative positions ("high skilled labor"), while the majority shifts towards worse-paid or even precarious forms of employment ("low skilled labor")[14]. This process of deskilling does not affect the unemployment rate, but there is still a need for political action.

Thirdly, as often, the narrative obscures the actual relationality and responsible actors. After all, artificial intelligence doesn't "take work away from people", and jobs don't disappear by magic. It's the decision-makers in corporations who decide that it might be more profitable to cut jobs, automate work processes, and/or relocate production. This development primarily depends on legal and economic framework conditions, corporate strategies, and political decisions regarding the role of AI in the societal transformation [86]. As long as the development of artificial intelligence is viewed as a force of nature and accompanied by fear scenarios like impending mass unemployment, it primarily benefits companies that produce or distribute AI products and corporations that benefit from fear-induced low wages. This narrative therefore hides the potential, and necessity, for societal shaping.

## 4 Conclusion and Outlook

By analyzing these AI narratives, it becomes evident that they often do not rest on the technical properties of these systems, frequently contain misconceptions about their context of use, and thus do not contribute to a fruitful and productive debate, sometimes even standing in its way. Their prominent presence in current discourses may be due to misunderstanding the technology, ignoring the societal embeddedness, or because such narratives just serve a commercial purpose.

AI is designed and used according to the purposes envisioned by companies, governments, and other organizations, sometimes with unintended side-effects. However, superficial Zeitgeist AI debates or dramatically sharpened contrasts à la "human versus machine" are misplaced, obscure the relevant power questions [54], and belong in the realm of science fiction.

Nevertheless, issues of complexity and controllability of these systems must be kept in view. There are long-standing methodological approaches and regulatory measures that could be taken [100], if there were the political will to do so. The lack of action is not least due to neoliberal societal images, coy "unknown new frontier" attitudes, powerful lobbying, innovation for its own sake, or even naive faith in technology. In this mix, questions of ethical and trustworthy AI are technically and philosophically interesting, but lag far behind the elaboration and practicability of discussions in computer science [87] the regulatory social sciences [35] and data protection theory [100], and sometimes ethics discussions even act as smokescreen for carrying on with the harmful business as usual [105]. The questions that arise when organizations deploy AI systems are indeed new but not fundamentally different from those that arise when organizations started to use cement, differential equations, or sniffer dogs for their purposes. However, a high degree of exceptionalism and depoliticization have always been key aspects of AI's self-narrative and of digitalization in general [80].

The deployment of AI technologies indeed transforms our societies, but not in the way often argued, whether utopian [3] or dystopian [52, 81]. Neither the salvation of humanity through a powerful and good AI, nor the impending extinction of humanity by an uncontrollable, thus evil super-intelligence, is to be expected. What is necessary right now is the collective negotiation of the actual effects and implications.

Actual consequences and dangers of these technologies, which must be the subject of societal debates, include the structural transformation of the labor market, the expansion of exploitative supply chains in data collection and analysis, especially in the Global South [83] the unilateral power and productivity increase of a few oligopolistic companies, the extension of personalized surveillance, the endless wave of AI-generated spam and fraud [18] the automated appropriation and commercialization of online cultural works, the neo-colonial (re)formation of global computation structures [71], the implications of military use of AI [1, 65, 124], the exponentially increasing energy consumption by AI systems [29], and not least the unreflective, ostensibly apolitical and hurtful deployment of such systems in the welfare state and other sensitive societal areas (social services, credit allocation, job application processes, legal proceedings, etc.) [10, 82]. It is crucial to keep in mind that technology design is always the design of (technologically shaped) social practices, but at the same time, the design is itself the result of social practices. Such socio-technical interdependence and complexity always have to be part of any technology reflection. And narratives heavily shape such interactions.

Luckily, the future is yet unwritten, and all these questions are up for discussion, so we as society can and should creatively and reflectively deliberate on how to use the interesting and powerful toolkit of new AI methods. Although many alternative theoretical and practical preliminary works exit, the application possibilities of AI and other digital technologies beyond profit and market have hardly been conceived and developed – from critical computer science [21, 113] and anarchist software design [63] to digital commoning [48] and non-authoritarian information systems [69], meaningfully supporting sustainability transformation [24, 26], reclaiming the means of computation [30] and even social justice and sustainability-aware tech utopias [98].

If we want to keep in sight the societal design alternatives of artificial intelligence and its diverse application possibilities, we need not only see through the recurring and misleading ELIZA moments fuelled by problematic narratives, but also consider whether a focus on massively complex AI systems appears globally beneficial, or whether such focus actually distracts from the real ingredients for a good life for all.

## Acknowledgments

I owe lots of gratitude to Paola Lopez, Jana Pannier and Stefan Ullrich for lengthy and insightful exchanges on the topic. This work was funded by the Federal Ministry of Education and Research of Germany (BMBF), grant no. 16DII131, and the Open Access Publication Fund of the Weizenbaum Institute for the Networked Society, Berlin.